\documentclass[a4paper]{jpconf}
\usepackage{graphicx}
\usepackage{wasysym}
\usepackage{color}
\usepackage{soul}

\begin{document}
\title{Progress on Brussels-Skyrme atomic mass models on a grid: stiff neutron matter equation of state}

\author{G. Grams$^1$, W. Ryssens$^1$, G. Scamps$^2$, S. Goriely$^1$ and N. Chamel$^1$.}
\address{$^{1}$Institut d’Astronomie et d’Astrophysique, Université Libre de Bruxelles, Campus de la Plaine CP 226, 1050 Brussels, Belgium}
\address{$^{2}$Department of Physics, University of Washington, Seattle, Washington 98195-1560, USA}

\ead{guilherme.grams@ulb.be}

\begin{abstract}
We report here the current developments on the Brussels-Skyrme-on-a-Grid (BSkG) atomic mass models. In comparison with our previous models, BSkG3 improves the infinite nuclear matter (INM) properties which opens its applications to neutron stars.
The results presented here show that BSkG3 preserve the excellent agreement with experimental nuclear masses and radii, together with fission barriers of actinides obtained by BSkG1 and BSkG2, while the nuclear matter properties are considerably improved.
\end{abstract}

\section{Introduction}

Experimentally inaccessible nuclear masses play a major role in our understanding of the r-process nucleosynthesis and the composition of neutron star (NS) crust.
The Brussels-Montréal Skyrme (BSk-series) \cite{Chamel09,Goriely16} and BSkG \cite{Scamps21,Ryssens22} mass models have been developed to this end, based on the Hartree-Fock-Bogoliubov (HFB) method with extended Skyrme forces, where the parameters are fitted to essentially all experimental nuclear masses.
Following the approach of BSkG1~\cite{Scamps21} and BSkG2~\cite{Ryssens22}, a new mass model was fitted. Details on the new model can be found in Ref.~\cite{Grams22}, in this contribution we present main features of the new model BSkG3.

\section{The BSkG3 mass model}

We obtain the nuclear mass from the mean-field energy of a HFB many-body state, complemented by perturbative corrections, as detailed in Refs.~\cite{Scamps21,Ryssens22}.
				We take advantage of features present on the BSk-series of mass models \cite{Chamel09,Goriely16} to improve the BSkG family. First, we introduce into the Skyrme force the $t_4$ and $t_5$ terms as in Ref.~\cite{Chamel09} which are density-dependent generalizations of the usual $t_1$ and $t_2$ terms. 
This modification allows us to produce a stiff equation of state (EoS) of neutron matter (NeutM) at high density and, at the same time, a low symmetry energy coefficient $J$ (see Sec.~\ref{sec:nucmatter}). In order to improve the pairing properties, we use a realistic treatment guided by ab-initio calculations~\cite{Cao06} as done in BSk30-32~\cite{Goriely16}. 

The fitting protocol follows the strategy of Ref.~\cite{Ryssens22} where a committee of Multi-Layer Neural Networks was used as an emulator to reduce the computational cost of the parameter adjustment (see Refs.~\cite{Scamps21,Grams22} for details). We fit the model parameters to essentially all experimental masses and to the five-point neutron mass-staggering $\Delta_n^{(5)}$. As a final step, we adjust the parameters of the collective correction to fission barriers so as to control the model properties at large deformations. Unlike BSkG2, in the present work we do not include a vibrational correction energy. 

\begin{figure}[h]
\includegraphics[width=37pc]{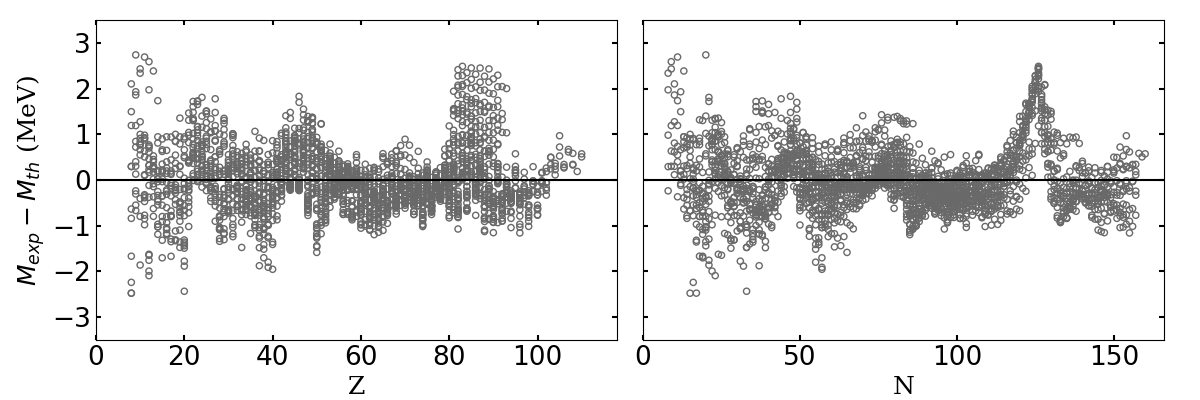}\hspace{2pc}%
\caption{\label{fig:masses}Mass differences between experimental masses\cite{AME2020} and BSkG3, as function of proton (left) and neutron number (right).}
\end{figure}

We display in Fig. \ref{fig:masses} the quality of the present fit by plotting the difference $M_{\mathrm{exp}}-M_{\mathrm{th}}$ against the number of protons $Z$ (left) and neutrons $N$ (right), with $M_{\rm exp/th}$ being the experimental/theoretical mass excess. 

Finally, the root-mean-square (rms) deviation $\sigma$ for the 2457 known masses of AME2020~\cite{AME2020} of nuclei with $Z$, $N \geq 8$, 884 measured charge radii $R_c$ \cite{Angeli13} and 45 reference values for primary ($E_I$) fission barrier heights~\cite{Capote09} for BSkG3 are $\sigma_{\rm BSkG3} (M) =  0.693$ MeV, $\sigma_{\rm BSkG3} (R_c) =  0.0250 $ fm and $\sigma_{\rm BSkG3} (E_I) =  0.38$ MeV, while with the previous BSkG1/BSkG2 we obtained $\sigma (M) = 0.741/0.678 $ MeV, $\sigma (R_c) =  0.0239/0.0274 $ fm and $\sigma (E_I) = 0.88/0.44 $ MeV.

\section{Nuclear matter properties}
\label{sec:nucmatter}

One goal of this work is to improve the description of INM properties calculated with the Skyrme interaction, mainly the stiffness of the EoS at high densities. The typical central density of NS cores computed with Skyrme models can reach 8-10 times saturation density, therefore models fitted to experimental masses (densities in nuclei are around or below saturation), without additional constraints, normally fail to describe the most massive NS due to a soft EoS. Thanks to the generalized density-dependent terms in the Skyrme force, we are able to include during the fit protocol a constraint on the NeutM EoS at high density, which ensures a stiff EoS while keeping a good quality of the fit to nuclear masses, as already shown in BSk30-32~\cite{Goriely16}.

\begin{figure}[h]
\includegraphics[width=37pc]{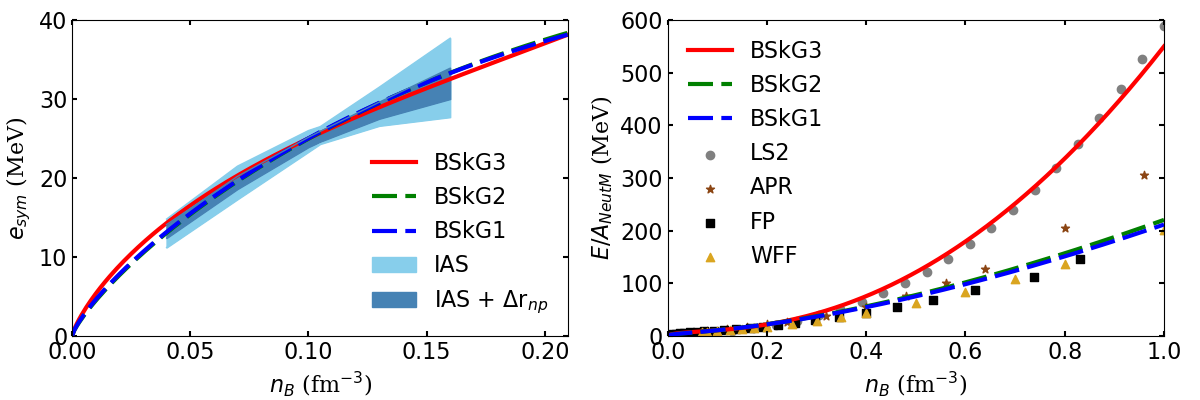}\hspace{2pc}%
\caption{\label{fig:nucmatt}{Left (right) panel shows the symmetry (neutron matter) energy predicted by BSkG1, BSkG2 and BSkG3 w.r.t baryon density. The BSkG models are compared with experimental data and {ab-initio} calculations from WFF\cite{WFF}, APR\cite{APR}, LS2\cite{LS2} and FP\cite{FP}.
}
}
\end{figure}

We present in Tab.~\ref{tab:nucmatter} the INM properties for the present work, for BSkG1\cite{Scamps21} and BSkG2\cite{Ryssens22}. 
In the left (right) panel of Fig.~\ref{fig:nucmatt} we compare the BSkG symmetry (NeutM) energy 
with experiments and realistic calculations. 
For the symmetry energy, our results reproduce well experimental constraints from isobaric analog state (IAS) and IAS + neutron skin, $\Delta r_{np}$, from Ref.~\cite{Danielewicz2014}. 
Regarding the NeutM energy BSkG3 presents a much stiffer behaviour at high densities when compared with our previous BSkG models and is comparable to the LS2 EoS.

\begin{table}[]
    \centering
\tabcolsep=0.12cm
\def\arraystretch{1.1}
    \begin{tabular}{cccccccccccc}
\hline\noalign{\smallskip}
Model &  $a_v$ & $n_0$ &  $J$&  $L$&  $K_{sym}$&  $K_v$&  $K^{'}$ &  $M^{*}_{s} / M$&  $M_v^*/M$&  $G_0$&  $G_{0}s^{'}$ \\
      &  (MeV) & fm$^{-3}$ &  (MeV)&  (MeV)&  (MeV)&  (MeV)& (MeV)&  &  & &  \\
\hline\noalign{\smallskip}
BSkG1     &  -16.09 & 0.158 &  32.0 & 51.7 & -156.4 & 237.8 & 376.7 & 0.86 & 0.77 & 0.35 & 0.98   \\
BSkG2     &  -16.08 & 0.158 &  32.0 & 53.0 & -150.6 & 237.5 & 376.3 & 0.86 & 0.77 & 0.36 & 0.98   \\
BSkG3  &  -16.09 & 0.158 &  31.0 & 51.4 & -42.4 & 244.6 & 296.9 & 0.86 & 0.69 & 0.05 & 0.98  \\
\noalign{\smallskip}\hline
    \end{tabular}
    \caption{Infinite nuclear matter properties for BSkG1~\cite{Scamps21}, BSkG2~\cite{Ryssens22} and BSkG3~\cite{Grams22} parameterizations. See Ref.~\cite{Goriely16} for the various definitions.}
    \label{tab:nucmatter}
\end{table}


\section{Conclusions}

The main improvements of the present work in comparison to previous BSkG models are: 
i) the stiff NeutM EoS at high densities, which avoids the collapse of NS and allows the description of heavy pulsars; 
ii) inclusion of a new pairing term as in Ref.~\cite{Goriely16}, which depends on the density gradient and allows the interaction to have a pairing gap that is fitted to realistic INM calculations \cite{Cao06} with self-energy corrections.
Both improvements, combined with our accurate description of nuclear structure properties, render the new BSkG3 model a tool of choice for applications in nuclear structure and astrophysics.

\ack
This work was supported by the Fonds de la Recherche Scientifique (F.R.S.-FNRS) and the Fonds Wetenschappelijk Onderzoek - Vlaanderen (FWO) under the EOS Projects nr O022818F and O000422F. Computational resources have been provided by the Consortium des Équipements de Calcul Intensif (CÉCI), funded by F.R.S.-FNRS under Grant No. 2.5020.11 and by the Walloon Region. GS is supported by U.S. Department of Energy, Office of Science,
Grant No.DE-AC05-00OR22725.

\bibliographystyle{iopart-num}       
\bibliography{ref.bib}   

\end{document}